\documentclass[11pt]{article}
\usepackage[T1]{fontenc} 
\usepackage{booktabs}
\usepackage{titlesec}
\titlelabel{\thetitle.\quad}
\usepackage{xcolor}
\usepackage{adjustbox}
\usepackage{amsmath, amsthm, amsfonts, amssymb}
\usepackage{graphicx}
\usepackage{setspace}
\usepackage{longtable}
\usepackage{enumitem}
\usepackage{breqn}
\usepackage{float}
\usepackage{siunitx}
\usepackage{lscape}
\usepackage{indentfirst}
\usepackage{pifont}
\usepackage[labelsep=period,justification=justified,singlelinecheck=false,font = footnotesize, labelfont=bf]{caption}
\usepackage{booktabs}
\usepackage{tabularx,ragged2e}
\usepackage{natbib}
\usepackage{rotating}
\usepackage{placeins}
\usepackage[colorlinks=true,linkcolor=black,urlcolor=blue]{hyperref}
\usepackage{subcaption}
\usepackage{caption} 

\usepackage{bbm}
\usepackage[margin=1in]{geometry}

\usepackage{enumerate}
\usepackage{array}
\usepackage[T1]{fontenc}
\usepackage{mathtools}

\DeclareCaptionLabelFormat{andtable}{#1~#2  \&  \tablename~\thetable}

\usepackage{fullpage, amsmath, amssymb, amsthm, bbm, color}
\usepackage{graphicx,caption,subcaption,placeins}
\usepackage{dsfont}

\usepackage{natbib}
\bibpunct{(}{)}{;}{a}{,}{,}

\newtheorem*{theorem*}{Theorem}

\usepackage{tikz}
\usetikzlibrary{positioning,chains,fit,shapes,calc}

\definecolor{myblue}{RGB}{80,80,160}
\definecolor{mygreen}{RGB}{80,160,80}

\begin{document}

\title{Racial Preferences at a Texas Medical School \\ \vspace{4mm} }

\author{David Puelz\thanks{Email: dpuelz@uaustin.org} \\ The University of Austin \\ \vspace{2mm}}

\date{\today}
\maketitle

\begin{abstract}
Whether and how race is used in selective admissions remains a central question in higher education and civil rights law. In \emph{Students for Fair Admissions v. Harvard} (2023), the Supreme Court held that race-based affirmative action in college admissions violates the Equal Protection Clause, purportedly ending the practice. This report examines admissions at a public medical school in the pre-\emph{SFFA} period. Using applicant-level data on over 11,000 applications to Texas Tech University Health Sciences Center Medical School for the 2021 and 2022 cycles, I relate admission decisions to academic merit (MCAT, GPA, science GPA), race, gender, and situational judgment (Casper) scores. Summary statistics, academic-index decompositions, and logistic regression models provide strong evidence of racial preferences: African American and Hispanic applicants are preferred relative to academically similar White and Asian applicants. Counterfactual and preference-removal analyses quantify the magnitude of these disparities. The findings document the kind of race-based preferences that \emph{SFFA} was meant to address and establish a baseline for assessing whether admissions practice changed after the decision.
\end{abstract}

\newpage
\setstretch{1.5}

\section{Introduction}

This report contains a statistical analysis of the admissions data for Texas Tech University Health Sciences Center Medical School during the 2021 and 2022 admission cycles.  First, I provide a summary of the data received and relevant statistics for the case.  Second, I consider how the data on each applicant is related to their chance of admission.  I employ standard statistical techniques to evaluate this relationship, and I show that there is strong evidence of admission penalty or preference related to race.  I conclude by summarizing the facts and statistical evidence presented.

\section{Data and Methods}

%

\subsection{Data}

I received two ``comma-separated values (csv)'' files containing applicant data for the 2021 and 2022 admission cycles.  The 2021 data file contains 5,990 applicants, and the 2022 data file contains 5,252 applicants.  For most of the following analysis, I aggregate across the two admission cycles and consider 5,990 + 5,252 = 11,242 applicants.  Across both cycles, there are seven variables provided.  They are:

\begin{itemize}[topsep=10pt, itemsep=0.75em, parsep=-7pt, label=\ding{213}]
	\item {\tt Status_Description}: The state of the application which takes one of sixteen values related to admission or rejection. I determine admission or rejection by collapsing this variable to those two levels, described below.
	\item {\tt MCAT}:  Applicant's score on the Medical College Admission Test which ranges from 472 to 528.
	\item {\tt GPA\_Science}: The grade point average of all undergraduate biology, chemistry, physics, and mathematics coursework.\footnote{See the \href{https://www.ttuhsc.edu/medicine/admissions/documents/Advising-Guide-2024-2025.pdf}{TTUHSC School of Medicine Guidebook}.}
	\item {\tt GPA\_Overall}: The overall undergraduate grade point average.
	\item {\tt Casper}: A normalized score assessing the applicant's ``situational judgement.''\footnote{See the \href{https://acuityinsights.app/casper/}{Casper website} for more details.  The skills assessed include ``collaboration, communication, empathy, equity, ethics, motivation, problem solving, professionalism, resilience and self-awareness.''}
	\item {\tt Gender}: A categorical variable taking on male, female, or decline to answer
	\item {\tt Ethnicity}: A categorical variable denoting the race of the applicant taking on one of 21 values.  I collapse this variable to seven unique categories that are described below.
\end{itemize}
I create a new variable from {\tt Status_Description} that takes a value of one if {\tt Status_Description} is equal to ``Matriculated,'' ``Withdrawn After Acceptance,'' ``Offer Declined,'' ``Deferred,'' or ``Admitted'' and a zero otherwise.  This new variable is an admission indicator and denotes whether or not an applicant was offered admission to the school. Across the two admission cycles, 607 applicants were admitted which represents a 5.41\% acceptance rate.

The {\tt Ethnicity} variable takes on one of 21 values.  I collapse these values into seven different racial groups/ethnicities--they are: ``White,'' ``African American,'' ``American Indian,'' ``Asian,'' ``Hawaiian/Pacific Islander,'' ``Hispanic,'' and ``Other.''  Please see Appendix \ref{app:race} for the mappings of the originally coded ethnicity values to the seven used in the analysis.

Please see Appendix \ref{app:samplestats} for summary statistics on the data set of admitted applicants (across both admission cycles).


\subsection{Methods}

I use standard statistical and econometric methods throughout the analysis.  First, I start with simple summary statistics, including estimated admission rates across different groups of applicants. Summary statistics are a useful starting point, but the question of racial penalty/preference is adequately probed only when applicant merit is also taken into account.  I do this by constructing a simple measure of merit a priori (an equally-weighted average of MCAT, GPA, and science GPA) which I call the ``academic index'' and examine the values of this measure across the applicant sample and within racial categories.  I find stark disparities in the academic index values across races.  Finally, I specify and fit a model of chance of admission (admission probability) related to these measures of merit and other applicant characteristics (race, gender, and score on the casper test).  This model is called a logistic regression and is a standard econometric approach for modeling a binary outcome like admission decision as a function of other observable data.  The estimates of this model describe Texas Tech's admission process parsimoniously, and I am able to investigate several dimensions of the process.  I conduct three analyses.  First, I generate predictions of admission for applicants of varying merit quality and examine the difference in predictions across race.  Second, I conduct a counterfactual analysis where I investigate how admission chances for an Asian male change had he been female and/or other races.  Third, I remove racial preferences and penalties to examine how the applicant pool make-up changes.  { \bf In all of these analyses, I observe that African American and Hispanic applicants are preferred at the expense of White and Asian applicants.  }

\section{Summary Statistics of Admission Rates}

I begin by examining the admission rates across racial categories.  Table \ref{tab:racerate} shows the admission rates and the size of the corresponding groups.  Whites are admitted at the highest rate and also represent the largest cohort of applicants.  Outside of Hawaiian/Pacific Islander with an admission rate of 0, African Americans and Hispanics have the lowest admission rates with relative large cohorts at roughly 1000 and 2000 applicants, respectively.  Asians represent the second largest cohort and have an admission rate at just over 5\%.  

I further divide the data into three classifications: (i) African American/Non-African American, (ii) White/Non-White, and (iii) Asian/Non-Asian.  The rates are shown in Tables \ref{tab:racerateblack}, \ref{tab:raceratewhite}, and \ref{tab:racerateasian}, respectively.  The difference in admission rates between African Americans and all other races is 2.65\% while Whites outpace all other races by 1.78\%.  The Asian/Non-Asian split is the closest with Non-Asians slightly ahead at 0.34\%.  A cursory analysis will rely only on these statistics to test the hypothesis of ``no racial discrimination.''  However, obfuscated in these rates are the academic qualifications of these applicants.  In further analyses, I show how taking merit and academic qualifications into account reveal a clear preference for African Americans and Hispanics at the expense of other races.      

\begin{table}[ht!] 
\centering
\begin{tabular}{lcc}
  \hline
race & rate (\%) & size \\ 
  \hline
White & 6.53 & 4152 \\ 
  African American & 3.00 & 1032 \\ 
  American Indian & 5.88 &   17 \\ 
  Asian & 5.17 & 3306 \\ 
  Hawaiian/Pacific Islander & 0.00 &    6 \\ 
  Hispanic & 4.37 & 1946 \\ 
  Other & 6.27 &  765 \\ 
   \hline
\end{tabular}
\caption{Admission rates by race. The statistics shown include the number of applicants in each subgroup in the final column.} 
\label{tab:racerate}
\end{table}

\begin{table}[ht!]
\centering
\begin{tabular}{lcc}
  \hline
race & rate (\%) & size \\ 
  \hline
African American & 3.00 &  1032 \\ 
  Non-African American & 5.65 & 10192 \\ 
   \hline
\end{tabular}
\caption{Admission rates by race, divided into African American and Non-African American. The statistics shown include the number of applicants in each subgroup in the final column.} 
\label{tab:racerateblack}
\end{table}

\begin{table}[ht!]
\centering
\begin{tabular}{lcc}
  \hline
race & rate (\%) & size \\ 
  \hline
    White & 6.53 & 4152 \\ 
Non-White & 4.75 & 7072 \\ 
   \hline
\end{tabular}
\caption{Admission rates by race, divided into White and Non-White. The statistics shown include the number of applicants in each subgroup in the final column.} 
\label{tab:raceratewhite}
\end{table}

\begin{table}[ht!]
\centering
\begin{tabular}{lcc}
  \hline
race & rate (\%) & size \\ 
  \hline
Asian & 5.17 & 3306 \\ 
  Non-Asian & 5.51 & 7918 \\ 
   \hline
\end{tabular}
\caption{Admission rates by race, divided into Asian and Non-Asian. The statistics shown include the number of applicants in each subgroup in the final column.} 
\label{tab:racerateasian}
\end{table}

\section{Admission Rates and Observable Characteristics}

The following section represents the bulk of my analysis of the data and evidence racial preferences and penalties.  I first investigate the question: To what extent are additional applicant characteristics, like academic ability, driving these differences in rate of admission?  

\subsection{Academic Index}

There are three variables in the data -- MCAT score, overall GPA, and science GPA -- that I use to define a simple academic index measure to evaluate the applicant's merit.  I a priori construct the academic index as an equally weighted combination of these three variables; namely, it is a weighted sum of the one-third MCAT, one-third overall GPA, and one-third science GPA. There are 509 out of 11201 applicants with missing data on at least one of these three variables (and thus the weighted index), so they are not considered for this analysis.  I also remove the Hawaiian/Pacific Islander and American Indian from the sample for ease of figure and table presentation given the sample size of 23.  Analysis with these races included will be provided upon request, but the conclusions will remain the same due to the very small cohort size relative to the entire applicant pool.

\begin{figure}[ht!]
\centering
	\includegraphics[scale=0.6]{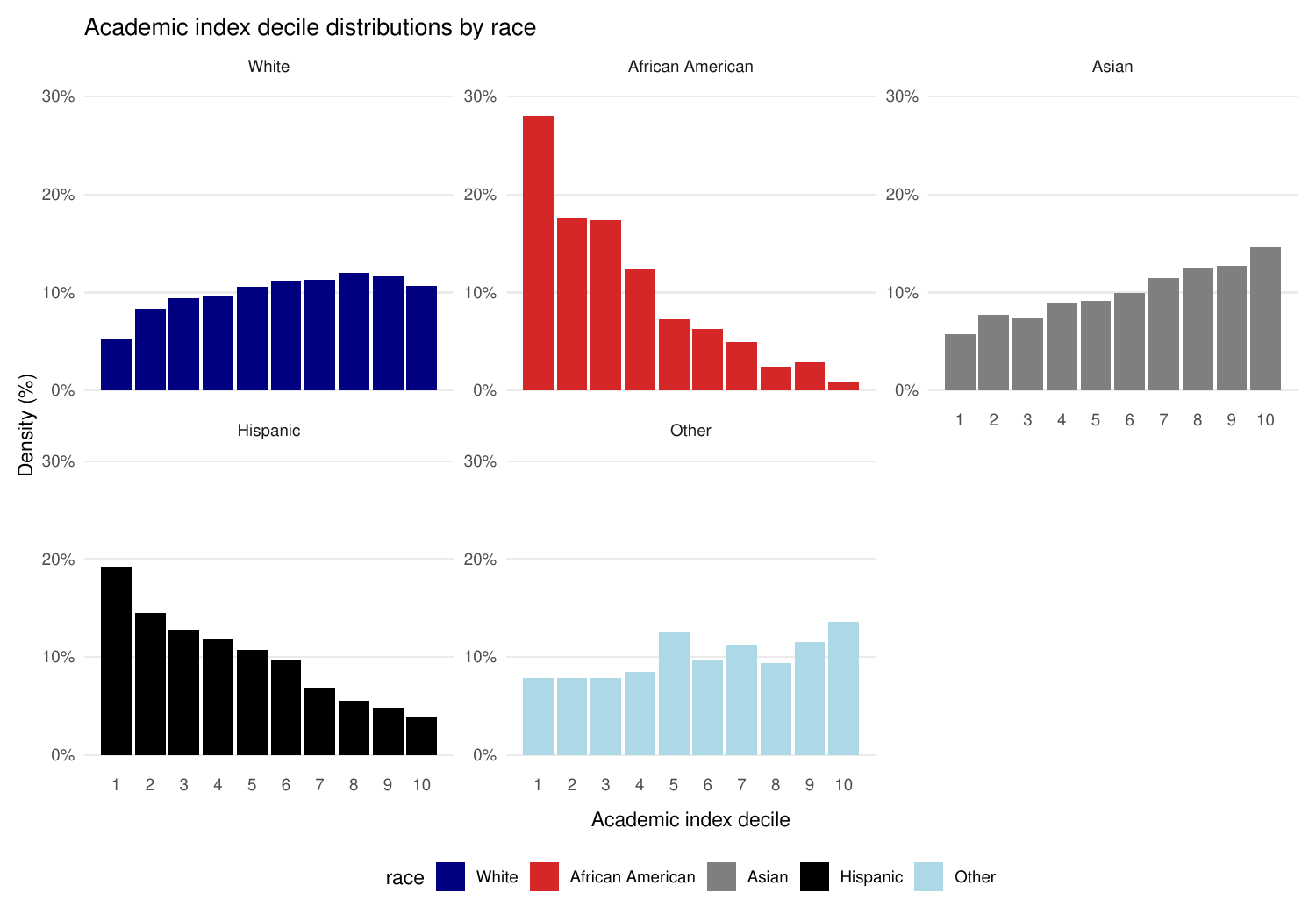}
	\caption{Distributions of the academic index deciles across racial categories.}
	\label{fig:AIrace}
\end{figure}

As a first step, I consider the distribution of the academic index across its deciles by race. By definition, the academic index is uniformly distributed across its own deciles, i.e., there are approximately the same number of applicants (10\% of the sample) located within each decile.  However, when separated out by race, the distributions tell a story of differing academic ability.  Specifically, there are different decile distributions across the racial categories.    This is shown in Figure \ref{fig:AIrace}.  Each panel in the figure represents a different race.  The bars display the percentage of applicants for a given race with an academic index value in the first decile, second decile, and so on.  For example, let's consider the African American panel.  The far left bar shows that almost 30\% of African American applicants have an academic index value in the bottom decile (where 90\% of the entire sample possess a higher index value).  In fact, over half of African American applicants have academic index values in the bottom three deciles.  This suggests that African Americans are ill-prepared academically relative to other races.  This is also the case for Hispanics, where nearly 20\% of applicants are in the bottom decile, and there is a downward trend for increasing deciles.

Whites, Asians, and Other all show slight increasing trends of decile proportions up to statistical fluctuation.  This means that while African Americans and Hispanics comprise the majority of the bottom decile academic index values, Whites and Asians comprise the majority of upper deciles.  This is most pronounced for Asians, where the vast majority are in the top half of academic ability.  Overall, what is clear from Figure \ref{fig:AIrace} is significant heterogeneity of academic ability across racial categories, and accounting for this variation will provide a clearer picture of the different admission rates across races.

%
%
%
%
%
%

\subsection{The Role of Race in Admissions}

In this section, I investigate the relationship between the academic index and admission decision.  Table \ref{tab:quantile_summary} below shows the admission rates for various subgroups.  The rows are the 10 deciles of the academic index, and the columns are the racial categories.\footnote{The heights of the bars in Figure \ref{fig:AIrace} are directly proportional to the subgroup sizes within each racial category.}  For example, the first decile (top row) are those applicants in the bottom 10\% of the academic index distribution, i.e., not academically strong.  Note that the admission rates across race for the first decile are all exactly 0, i.e., none of these applicants received an offer of admission.  Among the second decile applicants, African Americans and Hispanics each have nonzero admission rates while the others remain at zero.  Across the remaining eight deciles, African Americans and Hispanics often exceed the admission rates of similarly academically situated White and Asian counterparts.  For example, in the seventh decile, African Americans had 2x (14.9/7.5) the admission rate of Whites and 4.5x (14.9/3.3) the admission rate of Asians.  Hispanics similarly outperformed their White and Asian counterparts at this academic ability level.  For deciles three through eight, African Americans exceed their White/Asian counterparts in admission chances four out of six times and five out of six times, respectively.  The top two deciles have very few African Americans, hence the small or zero admission rates.  One final note is the Hispanic to Asian comparison.  Across the entire distribution of academic ability, Hispanics have equal (only at the bottom decile) or greater admission chances than Asians in nine out of ten deciles.
\begin{table}[ht!]
\centering
\begin{tabularx}{\textwidth}{>{\centering\arraybackslash}m{1cm}|>{\centering\arraybackslash}m{2.6cm}|
	>{\centering\arraybackslash}m{2.6cm}|>{\centering\arraybackslash}m{2.6cm}|>{\centering\arraybackslash}m{2.6cm}|>{\centering\arraybackslash}m{2.6cm}}
  \hline
Decile & White & African American & Asian & Hispanic & Other \\ 

 \hline
1 & 0.0\hfill (n=208) & 0.0\hfill (n=267) & 0.0\hfill (n=182) & 0.0\hfill (n=356) & 0.0\hfill (n=57) \\ 
  2 & 0.0\hfill (n=333) & 0.6\hfill (n=168) & 0.0\hfill (n=243) & 0.4\hfill (n=269) & 0.0\hfill (n=57) \\ 
  3 & 1.6\hfill (n=377) & 3.0\hfill (n=166) & 0.0\hfill (n=232) & 3.4\hfill (n=237) & 1.8\hfill (n=57) \\ 
  4 & 9.3\hfill (n=388) & 5.9\hfill (n=118) & 2.9\hfill (n=280) & 7.2\hfill (n=221) & 6.5\hfill (n=62) \\ 
  5 & 5.7\hfill (n=421) & 7.2\hfill (n=69) & 3.5\hfill (n=288) & 9.0\hfill (n=199) & 3.3\hfill (n=92) \\ 
  6 & 6.3\hfill (n=446) & 3.3\hfill (n=60) & 4.1\hfill (n=314) & 5.0\hfill (n=179) & 4.3\hfill (n=70) \\ 
  7 & 7.5\hfill (n=451) & 14.9\hfill (n=47) & 3.3\hfill (n=362) & 8.7\hfill (n=127) & 7.3\hfill (n=82) \\ 
  8 & 9.0\hfill (n=479) & 13.0\hfill (n=23) & 6.8\hfill (n=396) & 5.8\hfill (n=103) & 8.8\hfill (n=68) \\ 
  9 & 6.0\hfill (n=466) & 3.7\hfill (n=27) & 5.0\hfill (n=403) & 5.6\hfill (n=89) & 9.5\hfill (n=84) \\ 
  10 & 13.1\hfill (n=427) & 0.0\hfill (n=8) & 10.6\hfill (n=462) & 12.3\hfill (n=73) & 12.1\hfill (n=99) \\ 
 \hline
\end{tabularx}
\caption{Admission rates across deciles of the academic index and for each race.  Also included in parentheses are the sizes of each subgroup.} 
\label{tab:quantile_summary}
\end{table}

Why does this matter? Within each decile, the applicants have roughly the same academic ability.  So, if admission selection is entirely based on academic ability, the admission rates should be the same regardless of race up to statistical fluctuation.  Since this is not the case, there is either selection on race or on unmeasured variables.

A follow-up to this analysis is to consider the racial makeup of an admitted class if selection is purely based on academic ability.  To investigate, I run a simulation and present the results in Table \ref{tab:racesim} below.  The simulation is structured as follows:
\begin{itemize}[topsep=10pt, itemsep=0.75em, parsep=-7pt]
	\item Consider everyone at a given academic level (as measured by the academic index) equally likely for admission.
	\item Randomly sample 607 (the total number of admits across the observed sample) applicants and offer admission to this group.
	\item Observe the breakdown of this admitted group across racial categories.  
\end{itemize}
The simulation amounts to conducting a lottery for admission once you're deemed ``academically eligible'' as defined by your decile membership.  Importantly, there is no use of race or other applicant characteristics in this selection procedure.  For each academic level, the simulation is run 5000 times, and the results of the racial breakdown are averaged over the simulations.  The proportion of races sampled over different cohorts is shown in Table \ref{tab:racesim}.  The top row randomly samples applicants from the top 90\% of academic ability (leaving out the bottom 10\%).  The second row samples the top 80\% of academic ability.  The second-to-last row samples applicants from only the top decile of academic ability, i.e., only applicants who had an exceedingly high academic index value (greater than 90\% of the applicant pool).  The final row labeled ``observed proportions'' displays the observed racial breakdown of the admitted students in the sample.

\begin{table}[ht!]
\centering
\begin{tabularx}{\textwidth}{>{\centering\arraybackslash}m{2cm}|>{\centering\arraybackslash}m{2.4cm}|
	>{\centering\arraybackslash}m{2.4cm}|>{\centering\arraybackslash}m{2.4cm}|>{\centering\arraybackslash}m{2.4cm}|>{\centering\arraybackslash}m{2.4cm}}
  \hline
Sampling from top: & White & African American & Asian & Hispanic & Other \\ 
 \hline
90\% & 39.40 & 7.10 & 31.00 & 15.50 & 7.00 \\ 
  80\% & 40.40 & 6.00 & 32.00 & 14.40 & 7.20 \\ 
  70\% & 41.20 & 4.70 & 33.40 & 13.20 & 7.40 \\ 
  60\% & 41.90 & 3.60 & 34.70 & 12.00 & 7.70 \\ 
  50\% & 42.40 & 3.10 & 36.30 & 10.70 & 7.50 \\ 
  40\% & 42.60 & 2.50 & 38.00 & 9.20 & 7.80 \\ 
  30\% & 42.80 & 1.80 & 39.40 & 8.30 & 7.80 \\ 
  20\% & 41.70 & 1.60 & 40.50 & 7.60 & 8.60 \\ 
  10\% & 40.00 & 0.80 & 43.20 & 6.80 & 9.30 \\  
  \hline
    \hline
  Observed proportions & 44.70 & 5.10 & 28.20 & 14.00 & 7.90 \\
  \hline
\end{tabularx}
\caption{Racial breakdown for a random lottery simulation of admission based upon academic ability.  The top row samples the top 90\% of applicants based on the academic index, the next row samples the top 80\%, and so on.  The second-to-last row samples only the top decile of academic index applicants.  The final row displays the observed racial composition in the sample.} 
\label{tab:racesim}
\end{table}

African Americans and Hispanics are selected for admission more frequently in the observed proportions compared to the lottery, and this over-selection in the observed proportions comes with a severe penalty to Asians.  Notice that regardless of the academic ability of the cohort randomly sampled, the proportion of Asians generated from the lottery always exceed their observed proportion (28.2\%).  For example, consider sampling from the top decile only (second-to-last row).  This is realistic for a medical school seeking to admit highly qualified and academically strong applicants.  In this scenario, Asian applicants make up 43.2\% of the admits, a full 15 percentage points higher than their observed proportion of 28.2\%.  Across 607 admits, this is an additional 91 Asian admits over two application cycles.  Asians are so overrepresented in the top deciles of academic ability that they capture share of admits away from even Whites (this can be seen by Whites decreasing proportions in the top quintile and decile rows).  Note also that African Americans and Hispanics have sharply lower representation under the lottery--their share of admits is less than one-fifth and one-half that of their observed proportions, respectively.  This lottery demonstrates that race is strongly correlated with academic index--also seen in Figure \ref{fig:AIrace}--and that admission on academic ability alone would result in 15\% more Asians and far fewer African Americans and Hispanics than were actually admitted.

\subsection{Regressions}

Regression is an important tool for describing complex systems with known inputs and outputs.  The inputs and outputs are generally given by data, and a regression model specifies a relationship between those inputs and a single output of interest.  In our case, the inputs are observed applicant characteristics (MCAT, GPA, science GPA, etc.) and the output is the admission decision.  Since the admission decision is binary--an applicant is either accepted or rejected--I use a logistic regression model which specifies the odds of admission as a function of applicant characteristics.  Odds is defined as the ratio of the probability of admission to the probability of rejection.  

Table \ref{tab:coefficients} below shows the coefficient estimates for five regression models of increasing complexity.  The coefficient's sign denotes the partial effect of the characteristic on admission odds.  Positive denotes increasing values increases the odds, and negative denotes increasing values decrease the odds.  For example, model 1 estimates that a 1 point increase in an applicant's MCAT score will increase their odds of admission by 0.09.  Models 2 and 3 estimate a 1 point increase in GPA and science GPA leads to an increase of admission odds by roughly 4 and 3, respectively.  This directionality aligns with applicant selection based on academic ability.

Model 5 includes all academic characteristics as well as binary variables for race, gender, and the casper score.  The base category for this model are White males, meaning that the regression estimates for this group are incorporated into the intercept and all other estimates are offset relative to this subgroup.  Since this model uses all available applicant information provided, I utilize this description of the admission process for the remainder of the analysis.

\begin{table}[ht!]
\renewcommand{\arraystretch}{1.1} 
\begin{center}
\sisetup{parse-numbers=false, table-text-alignment=right}
\begin{tabular}{l S[table-format=5.5] S[table-format=5.5] S[table-format=5.5] S[table-format=5.5] S[table-format=5.5]}
\hline
 & {Model 1} & {Model 2} & {Model 3} & {Model 4} & {Model 5} \\
\hline
(Intercept)             & -47.00^{***} & -18.17^{***} & -14.30^{***} & -2.66^{***} & -45.96^{***} \\
                        & (3.12)       & (1.00)       & (0.74)       & (0.06)      & (3.85)       \\
MCAT                    & 0.09^{***}   &              &              &             & 0.07^{***}   \\
                        & (0.01)       &              &              &             & (0.01)       \\
GPA                     &              & 4.06^{***}   &              &             & 1.15         \\
                        &              & (0.26)       &              &             & (0.67)       \\
science GPA             &              &              & 3.09^{***}   &             & 1.32^{*}     \\
                        &              &              & (0.19)       &             & (0.52)       \\
African American    &              &              &              & -0.81^{***} & 0.30         \\
                        &              &              &              & (0.19)      & (0.21)       \\
Asian               &              &              &              & -0.25^{*}   & -0.49^{***}  \\
                        &              &              &              & (0.10)      & (0.11)       \\
Hispanic            &              &              &              & -0.42^{***} & 0.19         \\
                        &              &              &              & (0.13)      & (0.14)       \\
Other               &              &              &              & -0.04       & -0.00        \\
                        &              &              &              & (0.16)      & (0.18)       \\
Gender-NA 				&              &              &              &             & -11.30       \\
                        &              &              &              &             & (195.17)     \\
Female            &              &              &              &             & -0.09        \\
                        &              &              &              &             & (0.10)       \\
Casper score                  &              &              &              &             & -0.07        \\
                        &              &              &              &             & (0.05)       \\
\hline
AIC                     & 4110.26      & 4332.10      & 4325.84      & 4694.68     & 3657.45      \\
BIC                     & 4124.81      & 4346.74      & 4340.49      & 4731.30     & 3736.16      \\
Log Likelihood          & -2053.13     & -2164.05     & -2160.92     & -2342.34    & -1817.72     \\
Deviance                & 4106.26      & 4328.10      & 4321.84      & 4684.68     & 3635.45      \\
Num. obs.               & 10693        & 11200        & 11200        & 11201       & 9465         \\
\hline
\multicolumn{6}{l}{\scriptsize{$^{***}p<0.001$; $^{**}p<0.01$; $^{*}p<0.05$}}
\end{tabular}
\caption{Logistic regression models of admission probability.  Coefficients estimates are shown and standard errors are in parentheses.}
\label{tab:coefficients}
\end{center}
\end{table}

\begin{figure}[ht!]
\centering
	\includegraphics[scale=0.68]{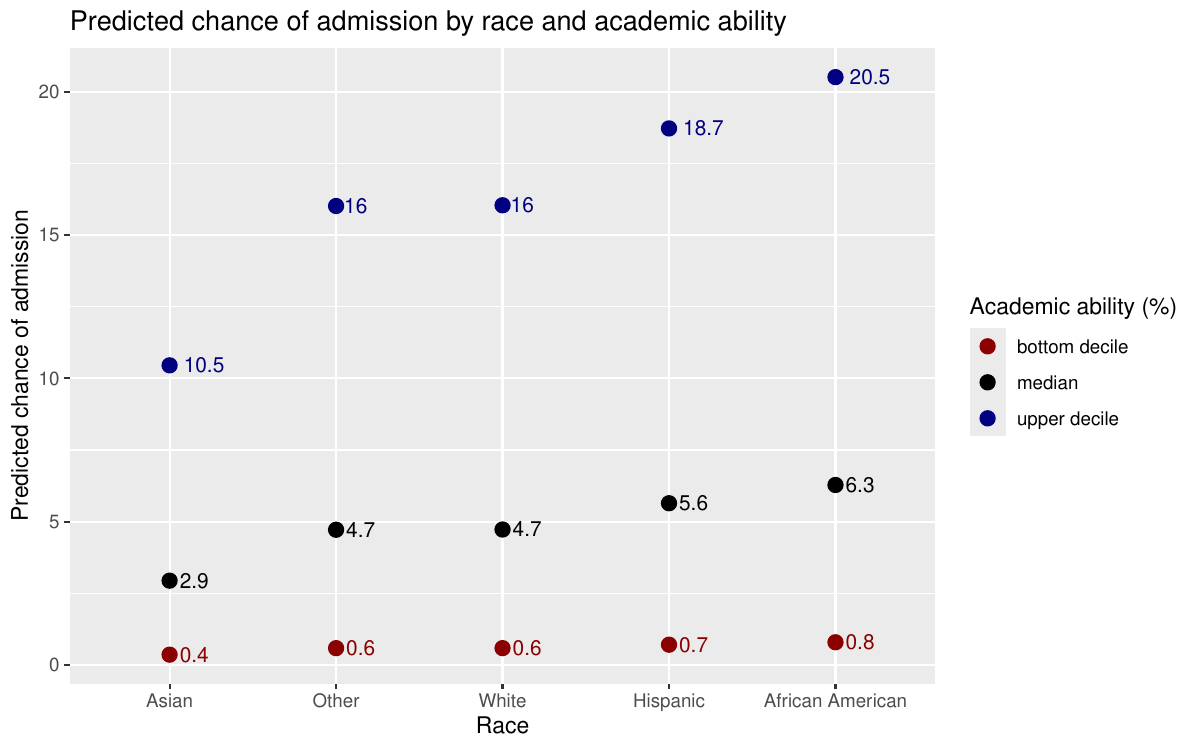}
	\caption{Predicted probability of admission for males with varying quantiles of MCAT, GPA, science GPA, and casper score using model 5. }
	\label{fig:raceprob}
\end{figure}

Once the regression model is estimated, I generate predictions of the chance of admission for any set of applicant characteristics. I conduct this analysis and show the results in Figure \ref{fig:raceprob}.  I consider a male applicant with an MCAT, GPA, science GPA, and casper score in the bottom 10\% (red), median (black), and top 10\% (blue).  I then vary race to observe how the predicted chance of admission changes.  The results show striking disparities among races despite the exact same observable characteristics.  For all three levels of academic ability, Asians and Whites are penalized relative to African Americans and Hispanics.  Assuming a top 10\% male applicant \textit{in all observable characteristics except for race}, an Asian's chance of admission is nearly half that of an African American.  A White's chance of admission is over four percentage points lower than an African American.

\vspace{3mm}
\noindent {\bf Counterfactual Analysis on Asian males}.  I further analyze the racial disparities reported in model 5 by setting applicant characteristics so that an Asian male has a probability of admission of 25\%.  I do this by fixing applicant characteristics at top 10\% values and adjust the intercept of the model so that the implied chance of admission is 25\%.  I then calculate the alternative--or counterfactual--probabilities of admission for varying races and genders.  They are shown in Table \ref{tab:asiancounter}.  For all alternative races and genders admission chances increase.  African American and Hispanic male applicants have the largest increases off of 25\% at 17 and 14 percentage points higher, respectively.  Across all races, females have lower chances compared to males.  These results are startling since all other observable characteristics are {\it exactly the same} except for gender and race and highlights disparities in the admission process.

\begin{table}[ht!]
\centering
\renewcommand{\arraystretch}{1.2} 
\setlength{\tabcolsep}{2pt}      
\begin{tabularx}{0.65\textwidth}{l>{\centering\arraybackslash}m{2.2cm}|>{\centering\arraybackslash}m{2.2cm}|>{\centering\arraybackslash}m{2.2cm}|>{\centering\arraybackslash}m{2.2cm}}
  \multicolumn{1}{c}{} & \multicolumn{4}{c}{Counterfactual Group} \\ 
  \cline{2-5}
  & White & African American & Hispanic & Other \\ 
  \cline{1-5}
  Male & 35.29 & 42.42 & 39.68 & 35.25 \\ 
  Female & 33.35 & 40.34 & 37.64 & 33.31 \\ 
  \cline{1-5}
\end{tabularx}
\caption{An Asian male with a 25\% chance of admission has the following admission chances if he were the following counterfactual genders and races.}
\label{tab:asiancounter}
\end{table}

\vspace{3mm}
\noindent {\bf Modulating Racial Preferences}. I now consider what would happen if racial penalties/preferences are removed from the admissions process.  I do this by setting coefficients on the different race variables in model 5 to zero.  Once coefficients are zeroed out, I recompute the model intercept to maintain the same number of admits across all racial categories, and this two-step process ensures the analysis accurately captures racial preference modulation.  Note that in using model 5, I require the data on all input variables to be recorded.  Removing the missing data leaves 9,465 applicants and 518 admits.

Table \ref{tab:modrace} displays the admit class composition under several changes in preferences.  In the ``No African American/Hispanic preference'' scenario, the number of admitted Whites grow by an additional 9 admits while African American and Hispanic admits drop by 6 and 10 respectively.  The ``No Asian penalty'' has the most significant effect.  Removing this penalty results in 49 additional Asian admits at the expense of all other races, especially Whites.  The second-to-last row zeroes out all race coefficients and hence removes all preferences and penalties.  Asians garner even more admits (55 more than observed) in this scenario, while African Americans and Hispanics lose 9 and 20, respectively.  

\begin{table}[ht]
\centering
\renewcommand{\arraystretch}{1.4} 
\setlength{\tabcolsep}{3pt}      
\begin{tabularx}{\textwidth}{l|>{\centering\arraybackslash}m{1.65cm}|
	>{\centering\arraybackslash}m{1.65cm}|>{\centering\arraybackslash}m{1.65cm}|>{\centering\arraybackslash}m{1.65cm}|>{\centering\arraybackslash}m{1.65cm}}
  \hline
  Racial preference change & White & African American & Asian & Hispanic & Other \\ 
  \hline
  No African American/Hispanic preference & 242 & 23 & 139 & 72 & 42 \\ 
  No Asian penalty & 203 & 25 & 183 & 72 & 35 \\ 
  No racial preferences whatsoever & 211 & 20 & 189 & 62 & 36 \\  
  \hline
  \hline
  Observed & 233 & 29 & 134 & 82 & 40 \\ 
  \hline
\end{tabularx}
\caption{Admit cohort composition when racial preferences/penalties are removed. The rows denote the changes in admission preference, and the numbers denote number of applicants of a given race admitted over the two-year application cycle. The bottom row displays the observed breakdown in the sample.}
\label{tab:modrace}
\end{table}

\begin{figure}[H]
\centering
	\includegraphics[scale=0.68]{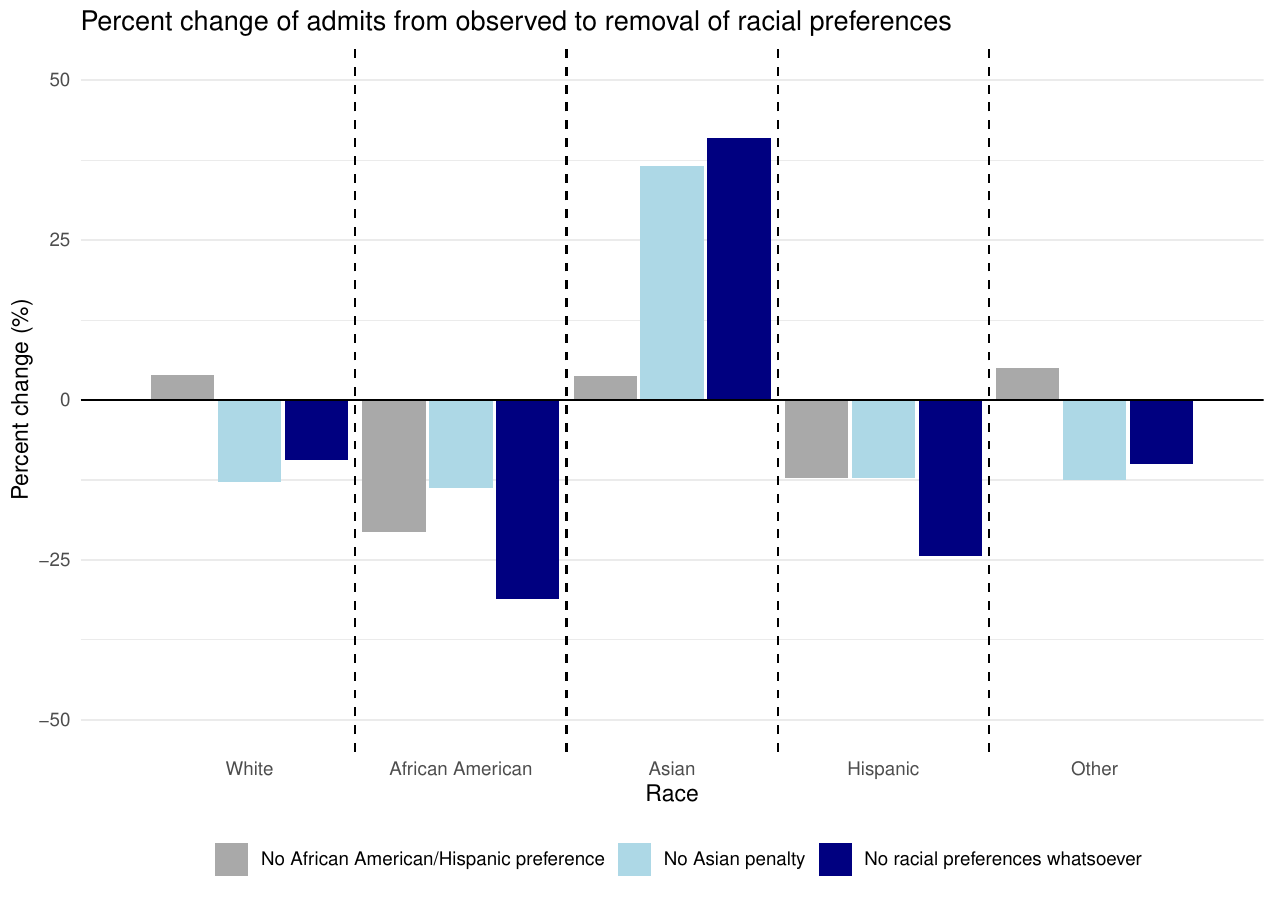}	
	\caption{Percentage change in admit cohort composition when racial preferences/penalties are removed.}
	\label{fig:modrace}
\end{figure}

I visualize these changes in Figure \ref{fig:modrace}.  The horizontal axis denotes race, and the colors display the three different changes to racial preferences/penalties.  A bar greater than zero means that race gains admit share when some or all preferences are removed.  A bar less than zero means that race loses admit share, implying they are preferentially treated during the admission process.  All of the Asian bars are greater than zero, while all of the African American and Hispanic bars are less than zero.  Whites experience an increase in admit share when African American/Hispanic preferences are removed.  This underscores a key finding that African Americans and Hispanics are preferentially treated at the expense of other races.  When these disparate impacts are removed, Whites and Asians stand to gain significantly.  

\section{Conclusion}

In this report, I conducted an analysis of two data files on admission to Texas Tech University Health Sciences Center Medical School.  First, I calculated the overall admission rates across races.  Second, I conditioned on observable characteristics of the applicants, including academic ability measured by MCAT score, GPA, and science GPA.  I showed through summary statistics, simulation, and regression modeling that African Americans and Hispanics are preferred for admission compared to academically identical White and Asian counterparts.

%
%
%
%
%

\newpage
\appendix

\section{Race Variable Definition}\label{app:race}

The original {\tt Ethnicity} variable contained 21 unique value across the 2021 and 2022 data sets.  This variable was collapsed to seven racial categories according to the following mapping:

\begin{table}[ht!]
\centering
\begin{tabular}{l|l}
\hline
Original value & New value \\
\hline
Black or African American  & African American  \\
Black or African American; Other & African American \\
American Indian or Alaska Native Other Hispanic, Latino, or of Spanish \\ origin; Hispanic, Latino, or of Spanish origin; Mexican/Chicano  & American Indian  \\
American Indian or Alaska Native; White Hispanic, Latino, or of Spanish \\ origin; Mexican/Chicano & American Indian \\
Hispanic, Latino, or of Spanish origin; Mexican/Chicano & Hispanic  \\
Puerto Rican; Hispanic, Latino, or of Spanish origin & Hispanic \\
Brazillian; Hispanic, Latino, or of Spanish origin & Hispanic \\
Dominican; Hispanic, Latino, or of Spanish origin & Hispanic \\
Brazil; Hispanic, Latino, or of Spanish origin & Hispanic \\
Multiple & Other \\
Unreported & Other \\
*blank*  & Other  \\
White/Caucasian & White \\
\hline
\end{tabular}
\end{table}
The remaining original value not displayed is ``Asian '' with a space after the ``n.''  This was mapped to simply ``Asian.''

\section{Admitted Sample Summary Statistics}\label{app:samplestats}

This appendix visualizes and displays data on the admitted sample across the two admissions cycles.  I include boxplots displaying the entire distributions and tables with summary statistics. 

\begin{figure}[H]
\centering
	\includegraphics[scale=0.6]{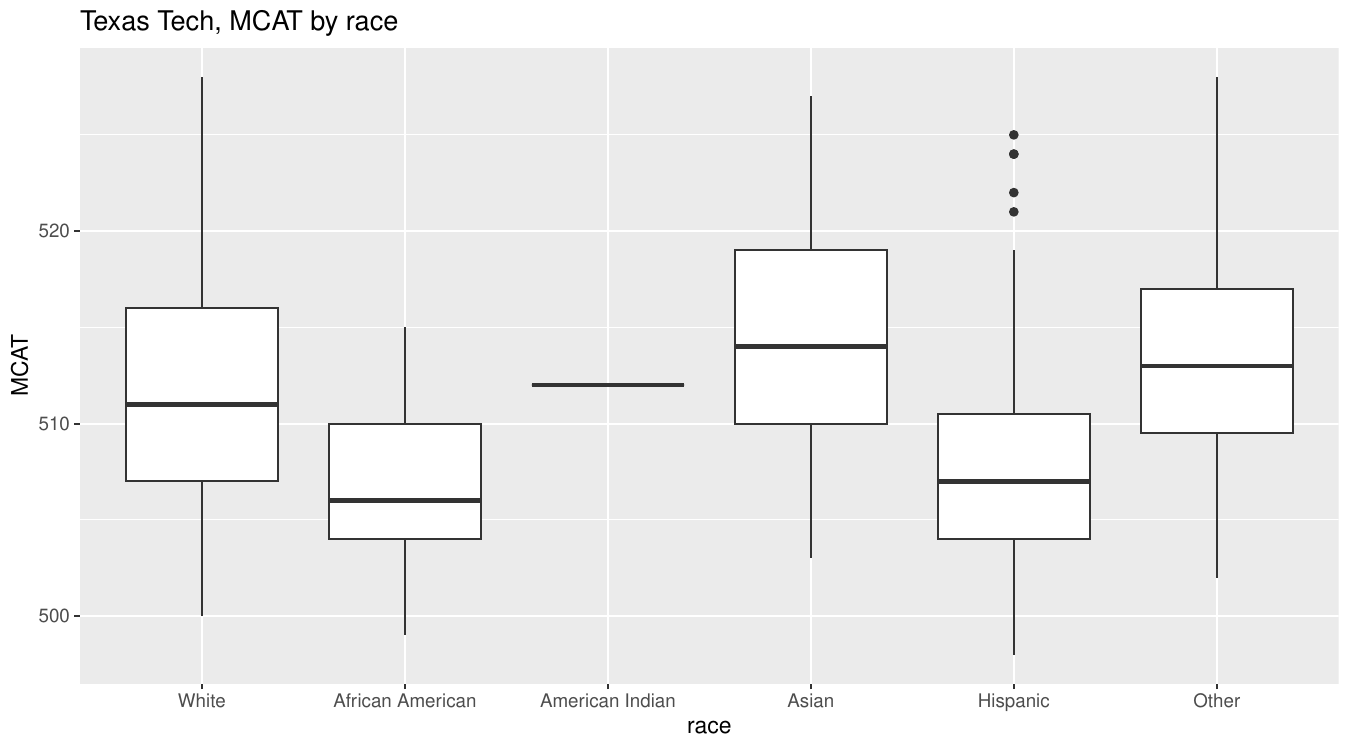}
	\caption{MCAT score by race. Boxes represent the inner-quartile-range (25th to 75th quantiles), and the solid black line represents the median.}
\end{figure}	

\begin{figure}[H]
\centering
	\includegraphics[scale=0.6]{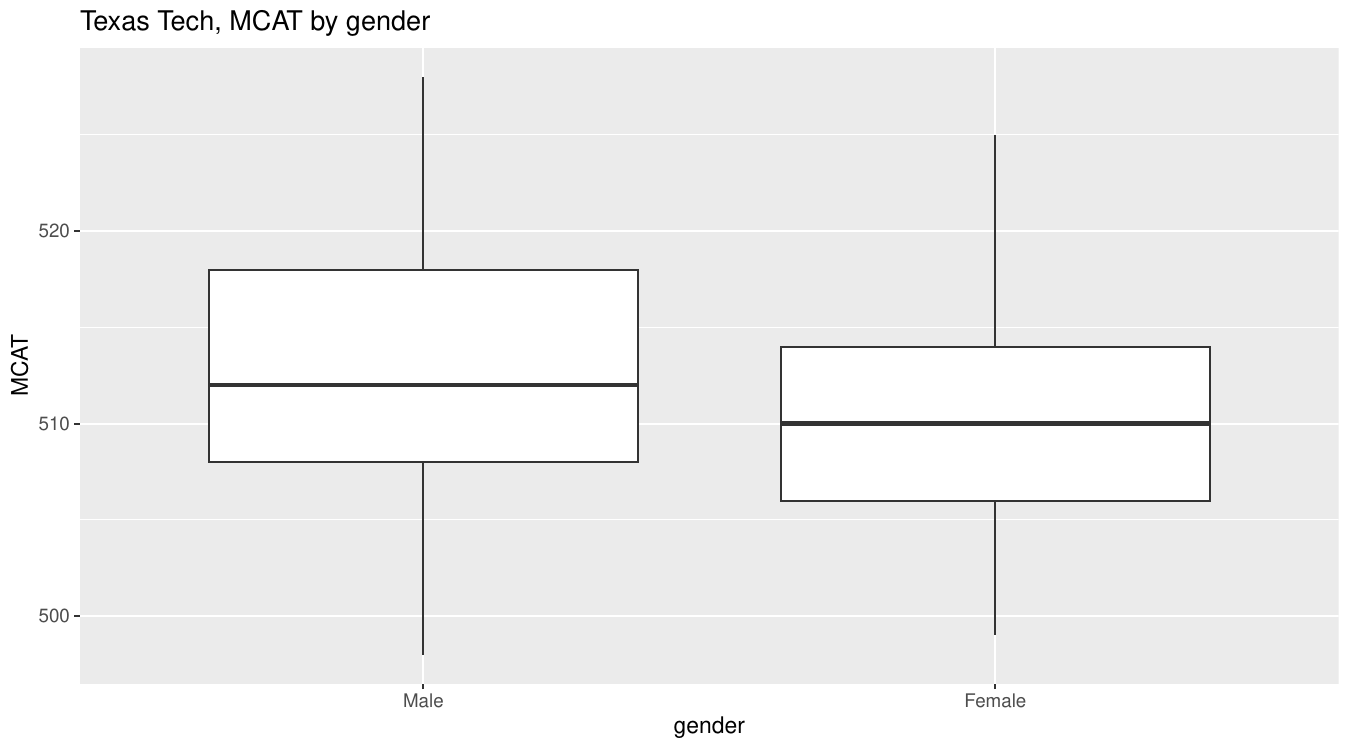}
	\caption{MCAT score by gender. Boxes represent the inner-quartile-range (25th to 75th quantiles), and the solid black line represents the median.}
\end{figure}	

\begin{figure}[H]
\centering
	\includegraphics[scale=0.6]{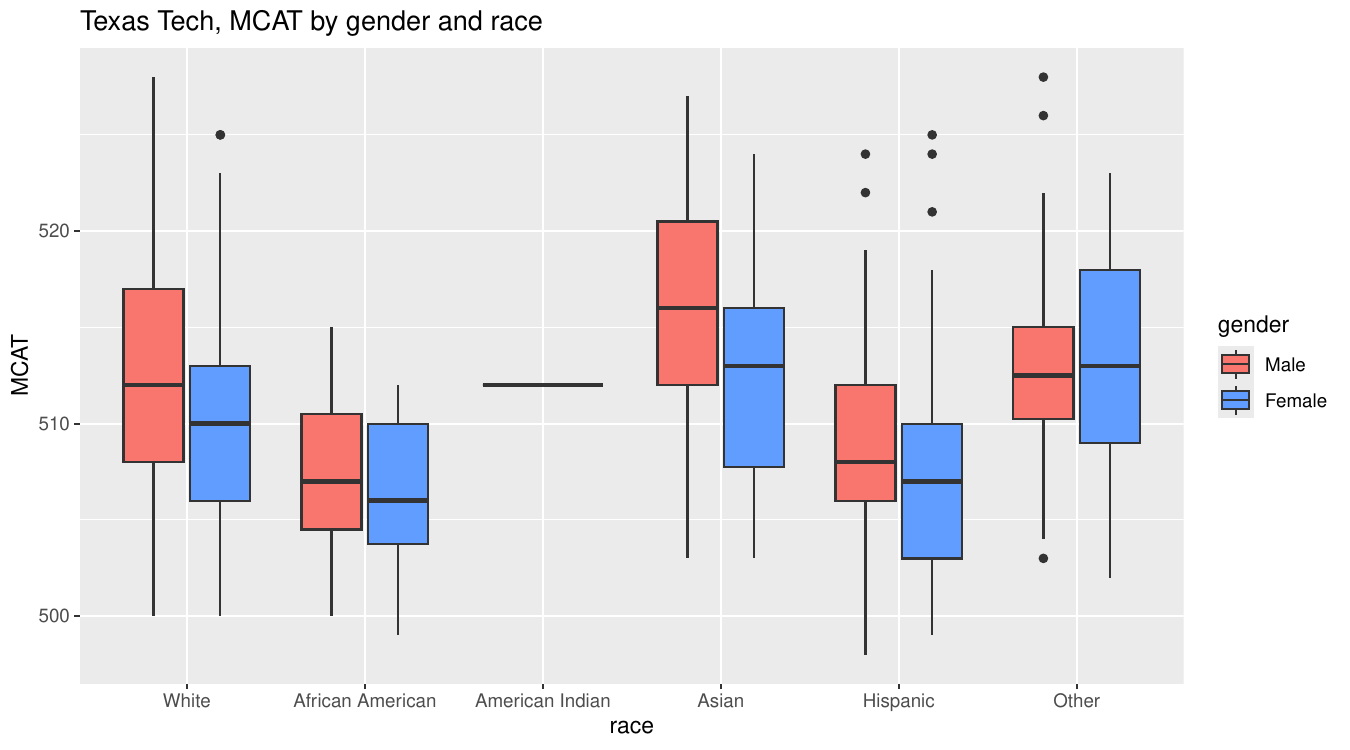}
	\caption{MCAT score by gender and race. Boxes represent the inner-quartile-range (25th to 75th quantiles), and the solid black line represents the median.}
\end{figure}

\begin{figure}[H]
\centering
	\includegraphics[scale=0.6]{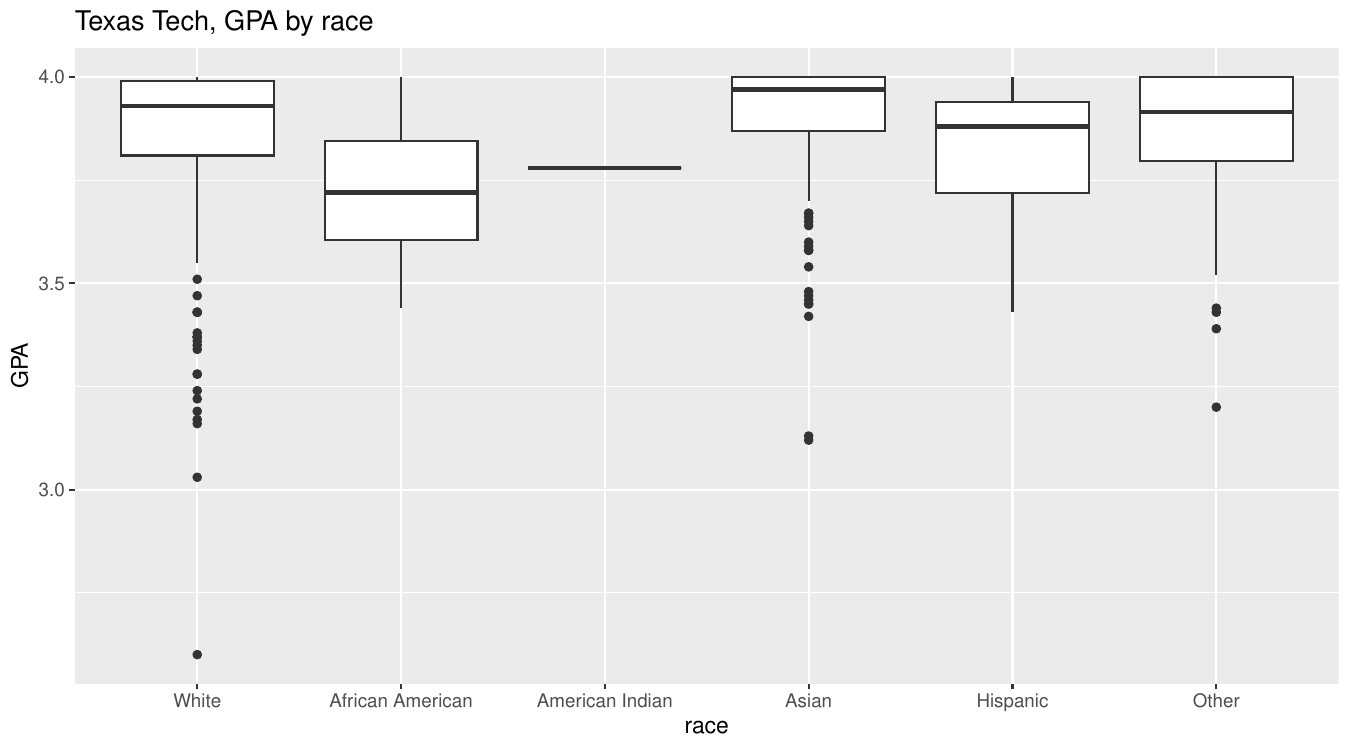}
	\caption{GPA by race. Boxes represent the inner-quartile-range (25th to 75th quantiles), and the solid black line represents the median.}
\end{figure}	

\begin{figure}[H]
\centering
	\includegraphics[scale=0.6]{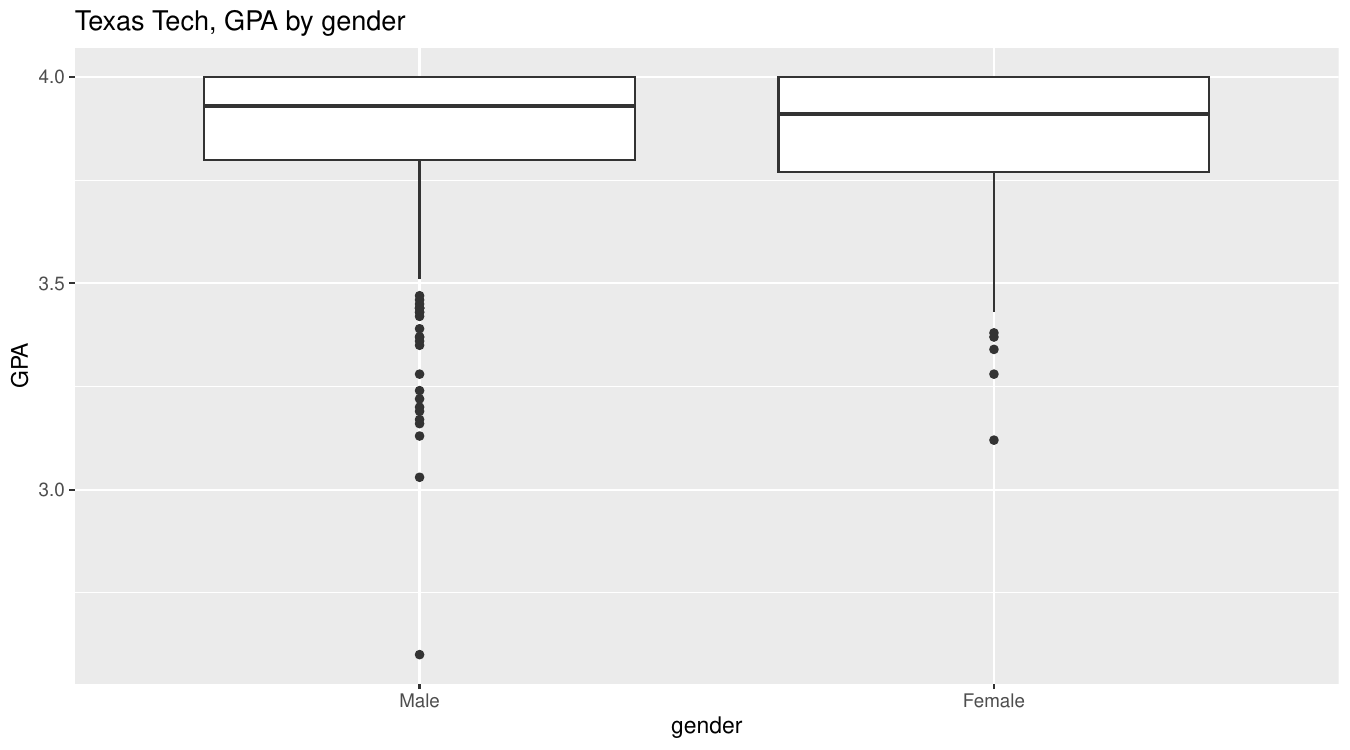}
	\caption{GPA by gender. Boxes represent the inner-quartile-range (25th to 75th quantiles), and the solid black line represents the median.}
\end{figure}	

\begin{figure}[H]
\centering
	\includegraphics[scale=0.6]{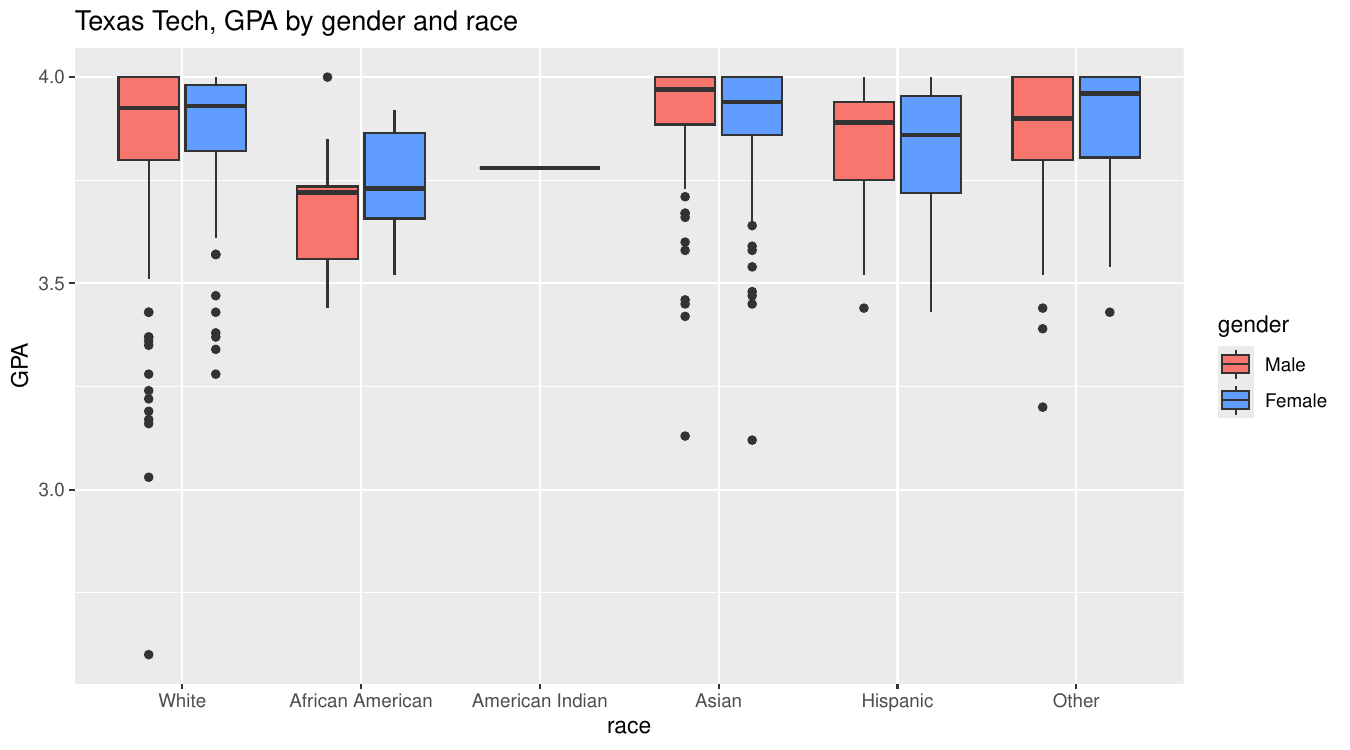}
	\caption{GPA by gender and race. Boxes represent the inner-quartile-range (25th to 75th quantiles), and the solid black line represents the median.}
\end{figure}

\begin{figure}[H]
\centering
	\includegraphics[scale=0.6]{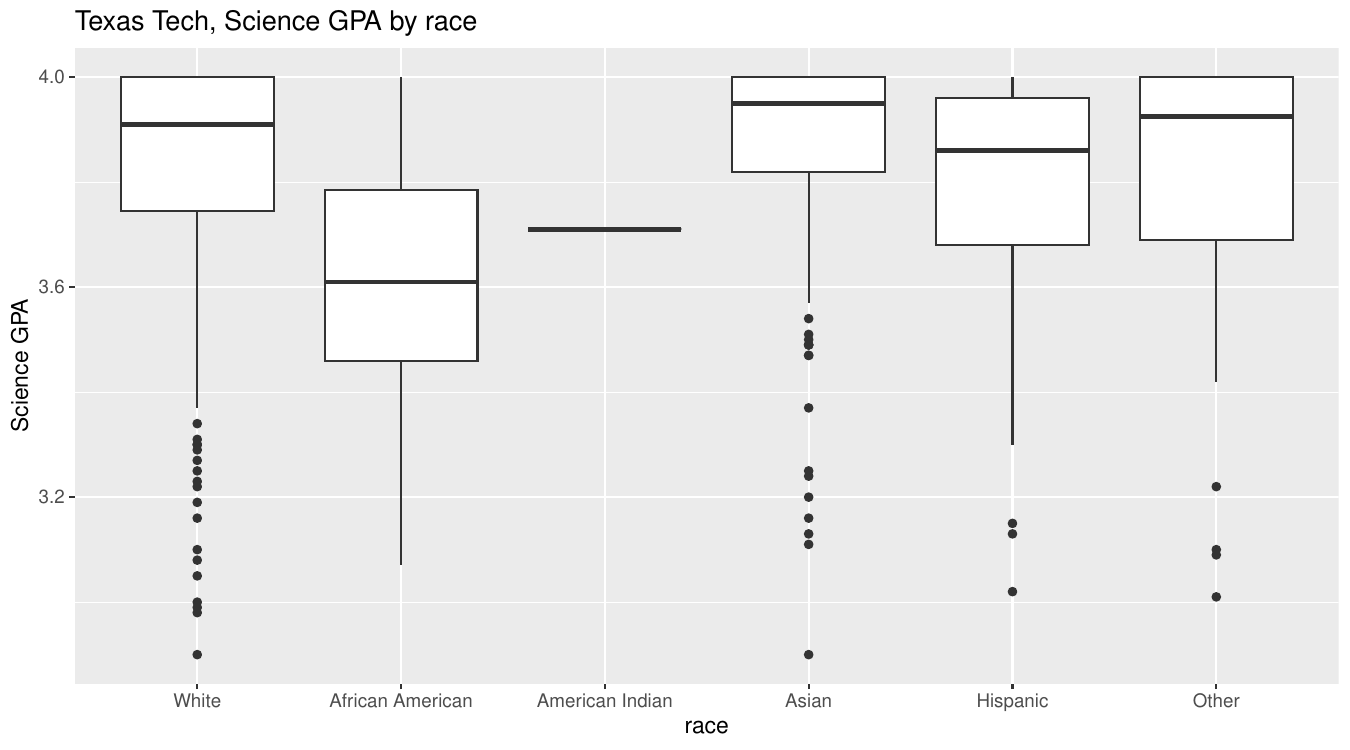}
	\caption{Science GPA by race. Boxes represent the inner-quartile-range (25th to 75th quantiles), and the solid black line represents the median.}
\end{figure}	

\begin{figure}[H]
\centering
	\includegraphics[scale=0.6]{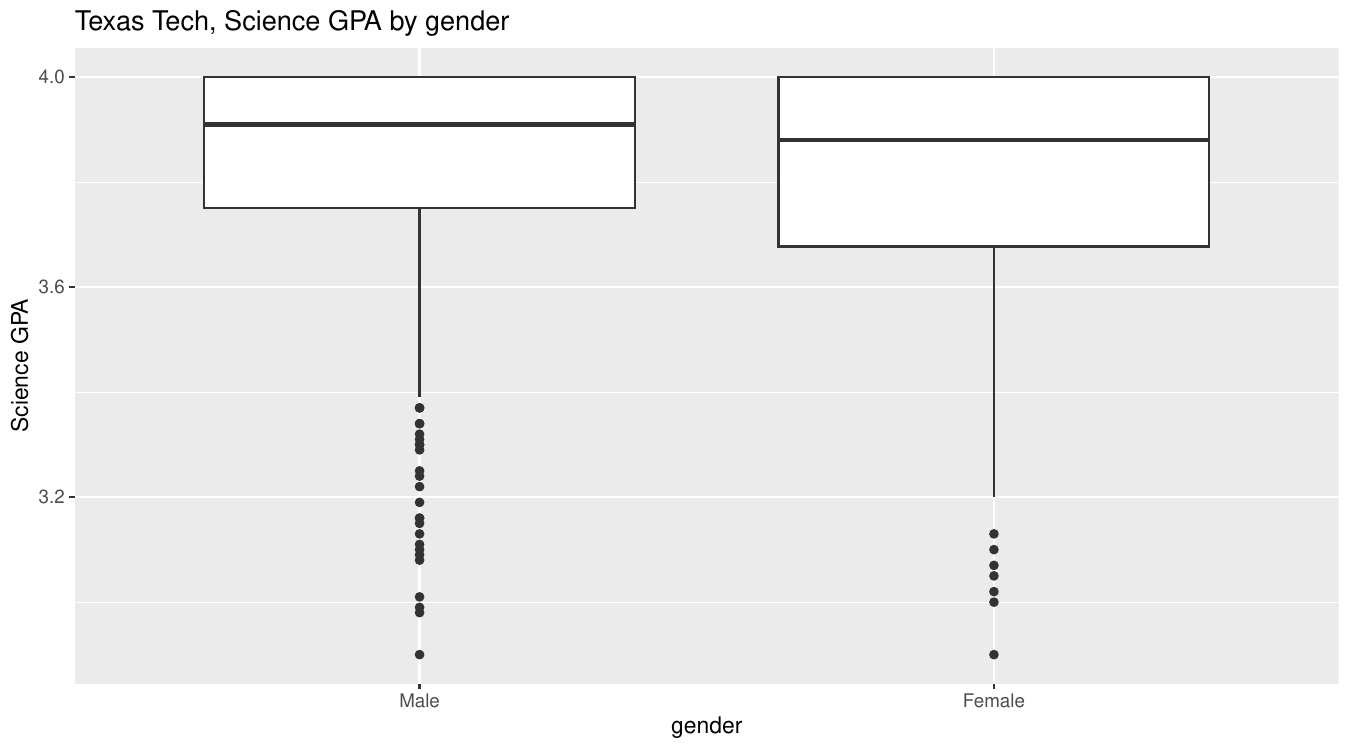}
	\caption{Science GPA by gender. Boxes represent the inner-quartile-range (25th to 75th quantiles), and the solid black line represents the median.}
\end{figure}	

\begin{figure}[H]
\centering
	\includegraphics[scale=0.6]{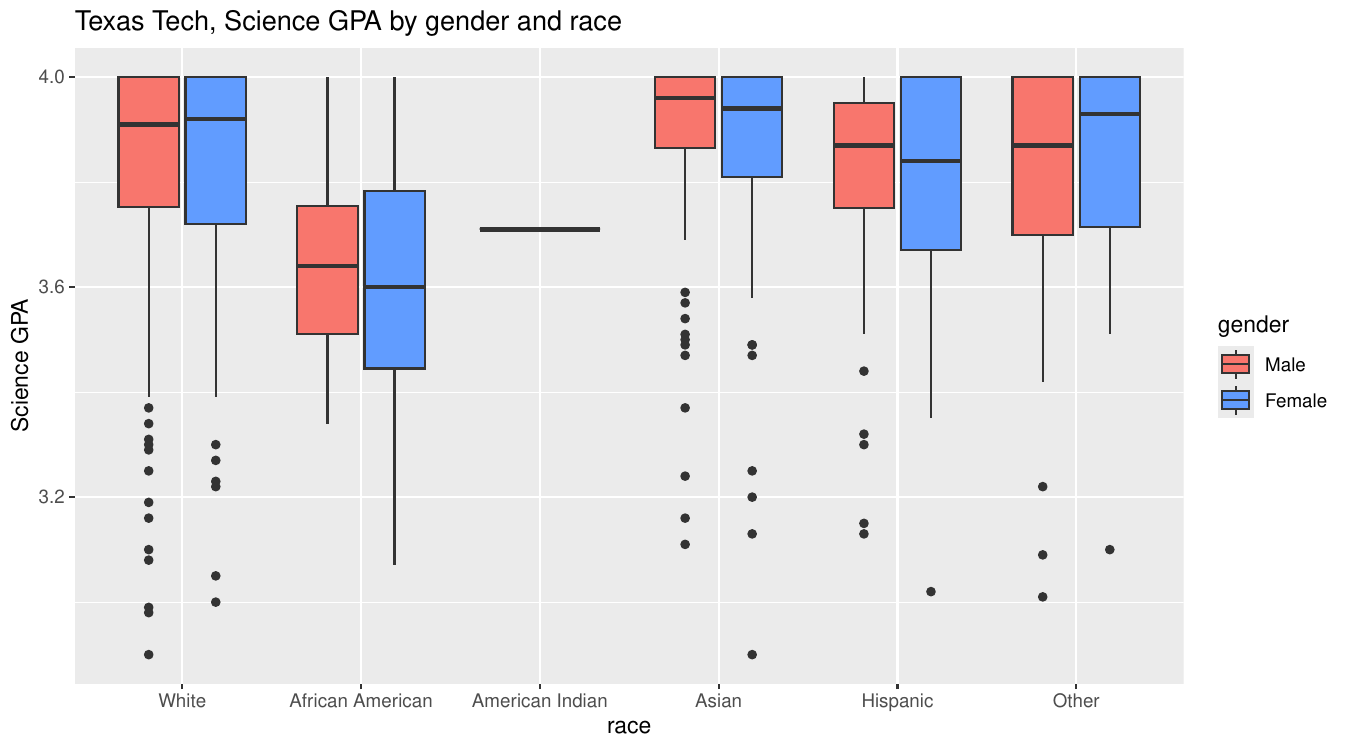}
	\caption{Science GPA by gender and race. Boxes represent the inner-quartile-range (25th to 75th quantiles), and the solid black line represents the median.}
\end{figure}

\begin{table}[ht]
\centering
\begin{tabular}{lllll}
  \hline
race & mean & median & sd & size \\ 
  \hline
White & 511.75 & 511 & 6.15 & 255 \\ 
  African American & 506.71 & 506 & 4.21 &  31 \\ 
  American Indian & 512.00 & 512 &  &   1 \\ 
  Asian & 514.51 & 514 & 6.09 & 139 \\ 
  Hispanic & 508.47 & 507 & 5.96 &  83 \\ 
  Other & 513.28 & 513 & 6.13 &  43 \\ 
   \hline
\end{tabular}
\caption{MCAT score summary statistics by race.} 
\end{table}
\begin{table}[ht]
\centering
\begin{tabular}{lllll}
  \hline
gender & mean & median & sd & size \\ 
  \hline
Male & 512.92 & 512 & 6.39 & 295 \\ 
  Female & 510.49 & 510 & 6.17 & 257 \\ 
   \hline
\end{tabular}
\caption{MCAT score summary statistics by gender.} 
\end{table}
\begin{table}[ht]
\centering
\begin{tabular}{llllll}
  \hline
race & gender & mean & median & sd & size \\ 
  \hline
White & Male & 512.67 & 512.0 & 6.10 & 147 \\ 
  African American & Male & 507.36 & 507.0 & 4.54 &  11 \\ 
  Asian & Male & 516.12 & 516.0 & 5.95 &  75 \\ 
  Hispanic & Male & 509.10 & 508.0 & 5.64 &  40 \\ 
  Other & Male & 513.41 & 512.5 & 6.22 &  22 \\ 
  White & Female & 510.50 & 510.0 & 6.02 & 108 \\ 
  African American & Female & 506.35 & 506.0 & 4.09 &  20 \\ 
  American Indian & Female & 512.00 & 512.0 &  &   1 \\ 
  Asian & Female & 512.62 & 513.0 & 5.74 &  64 \\ 
  Hispanic & Female & 507.88 & 507.0 & 6.25 &  43 \\ 
  Other & Female & 513.14 & 513.0 & 6.19 &  21 \\ 
   \hline
\end{tabular}
\caption{MCAT score summary statistics by race and gender.} 
\end{table}
\begin{table}[ht]
\centering
\begin{tabular}{lllll}
  \hline
race & mean & median & sd & size \\ 
  \hline
White & 3.85 & 3.93 & 0.21 & 271 \\ 
  African American & 3.71 & 3.72 & 0.14 &  31 \\ 
  American Indian & 3.78 & 3.78 &  &   1 \\ 
  Asian & 3.90 & 3.97 & 0.16 & 171 \\ 
  Hispanic & 3.83 & 3.88 & 0.16 &  85 \\ 
  Other & 3.86 & 3.92 & 0.19 &  48 \\ 
   \hline
\end{tabular}
\caption{GPA summary statistics by race.} 
\end{table}
\begin{table}[ht]
\centering
\begin{tabular}{lllll}
  \hline
gender & mean & median & sd & size \\ 
  \hline
Male & 3.85 & 3.93 & 0.21 & 323 \\ 
  Female & 3.86 & 3.91 & 0.16 & 284 \\ 
   \hline
\end{tabular}
\caption{GPA summary statistics by gender.} 
\end{table}
\begin{table}[ht]
\centering
\begin{tabular}{llllll}
  \hline
race & gender & mean & median & sd & size \\ 
  \hline
White & Male & 3.84 & 3.92 & 0.23 & 158 \\ 
  African American & Male & 3.68 & 3.72 & 0.16 &  11 \\ 
  Asian & Male & 3.90 & 3.97 & 0.16 &  88 \\ 
  Hispanic & Male & 3.83 & 3.89 & 0.16 &  41 \\ 
  Other & Male & 3.84 & 3.90 & 0.22 &  25 \\ 
  White & Female & 3.87 & 3.93 & 0.16 & 113 \\ 
  African American & Female & 3.73 & 3.73 & 0.13 &  20 \\ 
  American Indian & Female & 3.78 & 3.78 &  &   1 \\ 
  Asian & Female & 3.89 & 3.94 & 0.16 &  83 \\ 
  Hispanic & Female & 3.82 & 3.86 & 0.16 &  44 \\ 
  Other & Female & 3.88 & 3.96 & 0.16 &  23 \\ 
   \hline
\end{tabular}
\caption{GPA summary statistics by race and gender.} 
\end{table}
\begin{table}[ht]
\centering
\begin{tabular}{lllll}
  \hline
race & mean & median & sd & size \\ 
  \hline
White & 3.82 & 3.91 & 0.23 & 271 \\ 
  African American & 3.61 & 3.61 & 0.22 &  31 \\ 
  American Indian & 3.71 & 3.71 &  &   1 \\ 
  Asian & 3.86 & 3.95 & 0.21 & 171 \\ 
  Hispanic & 3.79 & 3.86 & 0.23 &  85 \\ 
  Other & 3.81 & 3.92 & 0.27 &  48 \\ 
   \hline
\end{tabular}
\caption{Science GPA summary statistics by race.} 
\end{table}
\begin{table}[ht]
\centering
\begin{tabular}{lllll}
  \hline
gender & mean & median & sd & size \\ 
  \hline
Male & 3.82 & 3.91 & 0.24 & 323 \\ 
  Female & 3.81 & 3.88 & 0.23 & 284 \\ 
   \hline
\end{tabular}
\caption{Science GPA summary statistics by gender.} 
\end{table}
\begin{table}[ht]
\centering
\begin{tabular}{llllll}
  \hline
race & gender & mean & median & sd & size \\ 
  \hline
White & Male & 3.82 & 3.91 & 0.24 & 158 \\ 
  African American & Male & 3.63 & 3.64 & 0.20 &  11 \\ 
  Asian & Male & 3.88 & 3.96 & 0.20 &  88 \\ 
  Hispanic & Male & 3.79 & 3.87 & 0.23 &  41 \\ 
  Other & Male & 3.78 & 3.87 & 0.30 &  25 \\ 
  White & Female & 3.82 & 3.92 & 0.23 & 113 \\ 
  African American & Female & 3.59 & 3.60 & 0.23 &  20 \\ 
  American Indian & Female & 3.71 & 3.71 &  &   1 \\ 
  Asian & Female & 3.85 & 3.94 & 0.22 &  83 \\ 
  Hispanic & Female & 3.78 & 3.84 & 0.23 &  44 \\ 
  Other & Female & 3.83 & 3.93 & 0.23 &  23 \\ 
   \hline
\end{tabular}
\caption{Science GPA summary statistics by race and gender.} 
\end{table}

\end{document}